\documentclass[11pt]{article}
\usepackage{microtype}
\usepackage[preprint]{acl}
\usepackage{times}
\usepackage{latexsym}
\usepackage{booktabs}
\usepackage{arydshln}
\usepackage{amsmath}
\usepackage{amsfonts}
\usepackage{xspace}
\usepackage{subcaption}
\usepackage{xcolor}
\usepackage{algorithm}
\usepackage{pifont}
\newcommand{\xmark}{\ding{55}}
\renewcommand{\checkmark}{\ding{51}}
\usepackage{natbib}
\definecolor{SeaGreen}{rgb}{0.18,0.55,0.34}
\usepackage[T1]{fontenc}
\usepackage[utf8]{inputenc}
\usepackage{inconsolata}
\usepackage{graphicx}
\usepackage{enumitem}
\usepackage{amssymb,amsthm}
\newtheorem{theorem}{Theorem}

\usepackage{algorithmic}
\usepackage{textcomp}
\usepackage{multirow}
\usepackage{url}

\title{AOI: Context-Aware Multi-Agent Operations with Dynamic Scheduling and Hierarchical Memory Compression}

\author{
  \textbf{Zishan Bai\textsuperscript{1}},
  \textbf{Hanxuan Chen\textsuperscript{2}$^*$},
  \textbf{Jiayi Gu\textsuperscript{3}$^*$},
  \textbf{Wenqian Weng\textsuperscript{4}},
  \textbf{Enze Ge\textsuperscript{5}},
  \textbf{Jiacheng Shi\textsuperscript{6}},\\
  \textbf{Yichao Zhang\textsuperscript{7}},
  \textbf{Zhimo Han\textsuperscript{8}},
  \textbf{Riyang Bao\textsuperscript{9}},
  \textbf{Xinyuan Song\textsuperscript{9}},
  \textbf{Jacqueline Pang\textsuperscript{10}},
  \textbf{Junfeng Hao\textsuperscript{11}$^\dagger$},
\\
  \textsuperscript{1}Columbia University \quad
  \textsuperscript{2}Hunan University \quad
  \textsuperscript{3}Central University of Finance and Economics \\
  \textsuperscript{4}Wayne State University \quad
  \textsuperscript{5}AI Agent Lab, Vokram Group \quad
  \textsuperscript{6}College of William and Mary \\
  \textsuperscript{7}University of Texas \quad
  \textsuperscript{8}Zhengzhou University of Light Industry \quad
  \textsuperscript{9}Emory University \\
  \textsuperscript{10}Cornell University \quad
  \textsuperscript{11}Department of Nephrology, Affiliated Hospital of Guangdong Medical University \\
  \small{
    \textbf{$^*$ Equal contribution.} \quad
    \textbf{$^\dagger$ Corresponding author:}
    \href{mailto:ygzhjf85@gmail.com}{ygzhjf85@gmail.com}
  }
}

\begin{document}

\maketitle
\begin{abstract}
Cloud-native systems have made operational work both more powerful and harder to automate: incidents unfold across microservices, logs and metrics arrive faster than operators can inspect them, and recovery actions must be coordinated without losing the causal context that makes them safe. We present AOI (AI-Oriented Operations), a context-aware multi-agent framework for autonomous IT operations. AOI separates operational responsibility across an Observer, a read-only Probe, and a guarded Executor, and connects them through dynamic scheduling and a hierarchical memory system with LLM-based context compression. This design turns long-running incident response into an iterative loop of observation, evidence gathering, safe intervention, and memory update. Across AIOpsLab simulations and real-world Loghub-derived scenarios, AOI improves task success to 94.2\%, reduces mean time to resolution by 34.4\% relative to the strongest baseline, and compresses operational context by 72.4\% while preserving 92.8\% of diagnostic information. Ablations show that these gains come from the combination of agent specialization, adaptive scheduling, and memory-aware compression rather than from any single module alone. The results suggest that autonomous operations systems can move beyond alert classification toward reliable, context-preserving recovery in complex infrastructure.
\end{abstract}
\section{Introduction}
Modern IT infrastructures have become distributed, elastic, and deeply interdependent systems built from microservices, containerized workloads, and dynamically orchestrated resources \cite{dragoni2017microservices,gannon2017cloud}. 
This architecture improves deployment velocity, but it also changes the nature of operational failure: a local slowdown, stale configuration, or noisy dependency can propagate through many services before the root cause is visible \cite{soldani2018pains,taibi2019microservices}. 
At the same time, production systems generate massive streams of logs, metrics, traces, and alerts, often at a scale that makes manual inspection impractical during an incident \cite{xu2009detecting,brewer2016sre}. 
The central challenge is therefore no longer only detecting anomalies; an operational agent must preserve context, decide what evidence to collect next, and act safely under partial information.

\begin{figure*}[t!]
\centering
\includegraphics[width=0.9\linewidth]{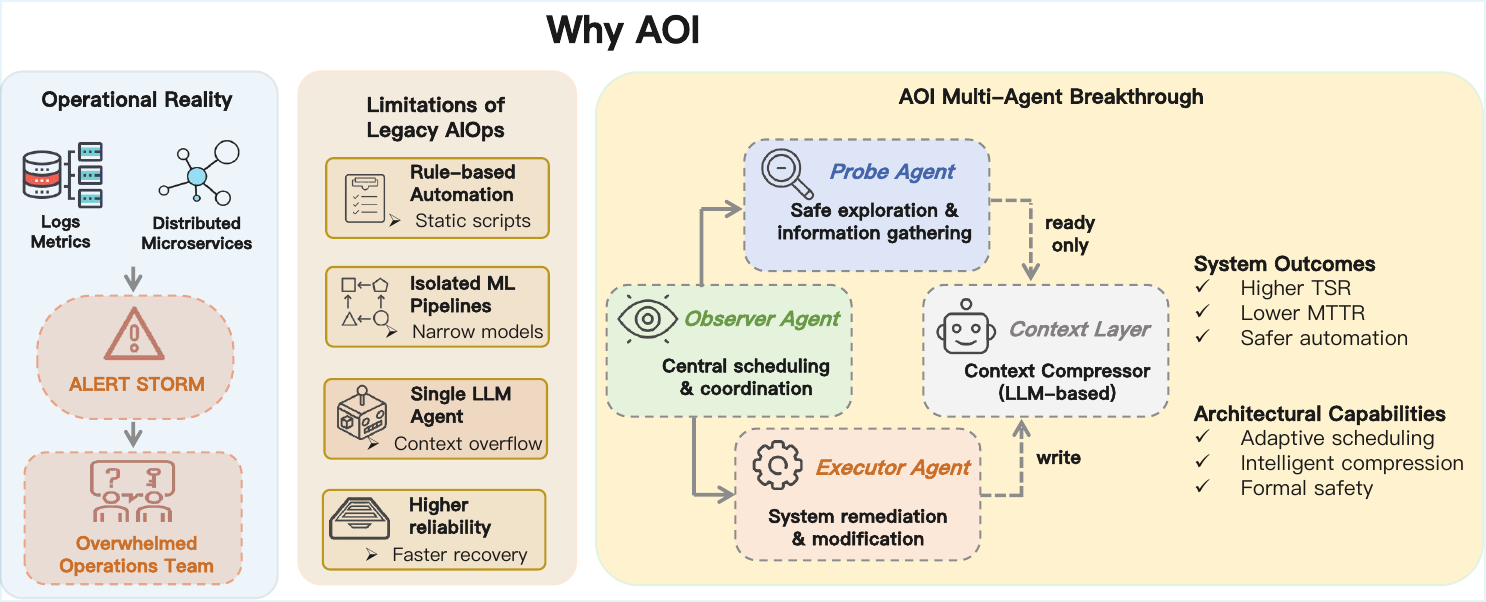}
\caption{Motivation for AOI. Modern operations require an agentic loop that maintains incident context across observation, diagnosis, intervention, and recovery. AOI addresses the resulting information overload by combining specialized agents, dynamic scheduling, and memory-aware context compression.}
\label{fig:aoi_architecture}
\end{figure*}

Automated operations research has made progress along three main lines. Expert systems provide transparent rules but require extensive manual engineering and degrade as infrastructure changes \cite{xu2017belief,muslewski2022expert,sarazin2021expert}. Machine-learning AIOps systems improve anomaly detection and log analysis, yet they are usually optimized for isolated prediction tasks rather than end-to-end diagnosis and recovery \cite{du2017deeplog,meng2019loganomaly,dang2019aiops,he2021survey}. Multi-agent automation introduces coordination and decentralization, but existing designs rarely maintain a shared long-horizon operational memory or adapt their information-gathering strategy as an incident unfolds \cite{wooldridge2009introduction,seitz2021automation,d2021designing,pulikottil2021multi}. Recent LLM-agent and tool-use benchmarks further show that agents can reason, call tools, and interact with environments, but they also expose persistent weaknesses in long-horizon planning, API reliability, and execution-grounded evaluation \cite{yao2023react,shinn2023reflexion,yao2023tree,liu2023agentbench,zhou2023webarena,xie2024osworld,jimenez2023swebench,qin2023toolllm,li2023apibank,patil2023gorilla,guo2024stabletoolbench,mialon2023gaia}.

These threads point to four requirements for autonomous IT operations. First, the system must filter high-volume telemetry without discarding the evidence needed for root-cause analysis \cite{Loghub,liu2024lost}. Second, it must coordinate probing and execution rather than treating diagnosis and remediation as separate pipelines \cite{AIOpsLab,wu2023autogen}. Third, it must preserve context across a long incident trajectory, where early observations often become decisive only after later evidence arrives \cite{hu2025hiagent,lewis2020retrieval}. Finally, it must keep system-modifying actions behind explicit safety checks, because an incorrect repair can be more damaging than a delayed one \cite{TLAPLUS,brewer2016sre}.

To address these challenges, we present AOI (AI-Oriented Operations), a context-aware multi-agent framework for autonomous IT operations (Figure~\ref{fig:aoi_architecture}).
AOI decomposes incident response into specialized roles, schedules evidence collection and intervention dynamically, and compresses operational context into a reusable memory state.
Table~\ref{tab:related_work_comparison} positions AOI against representative AIOps, multi-agent, and context-compression approaches.
Our main contributions are:
\begin{figure*}[t!]
\centering
\includegraphics[width=0.85\textwidth]{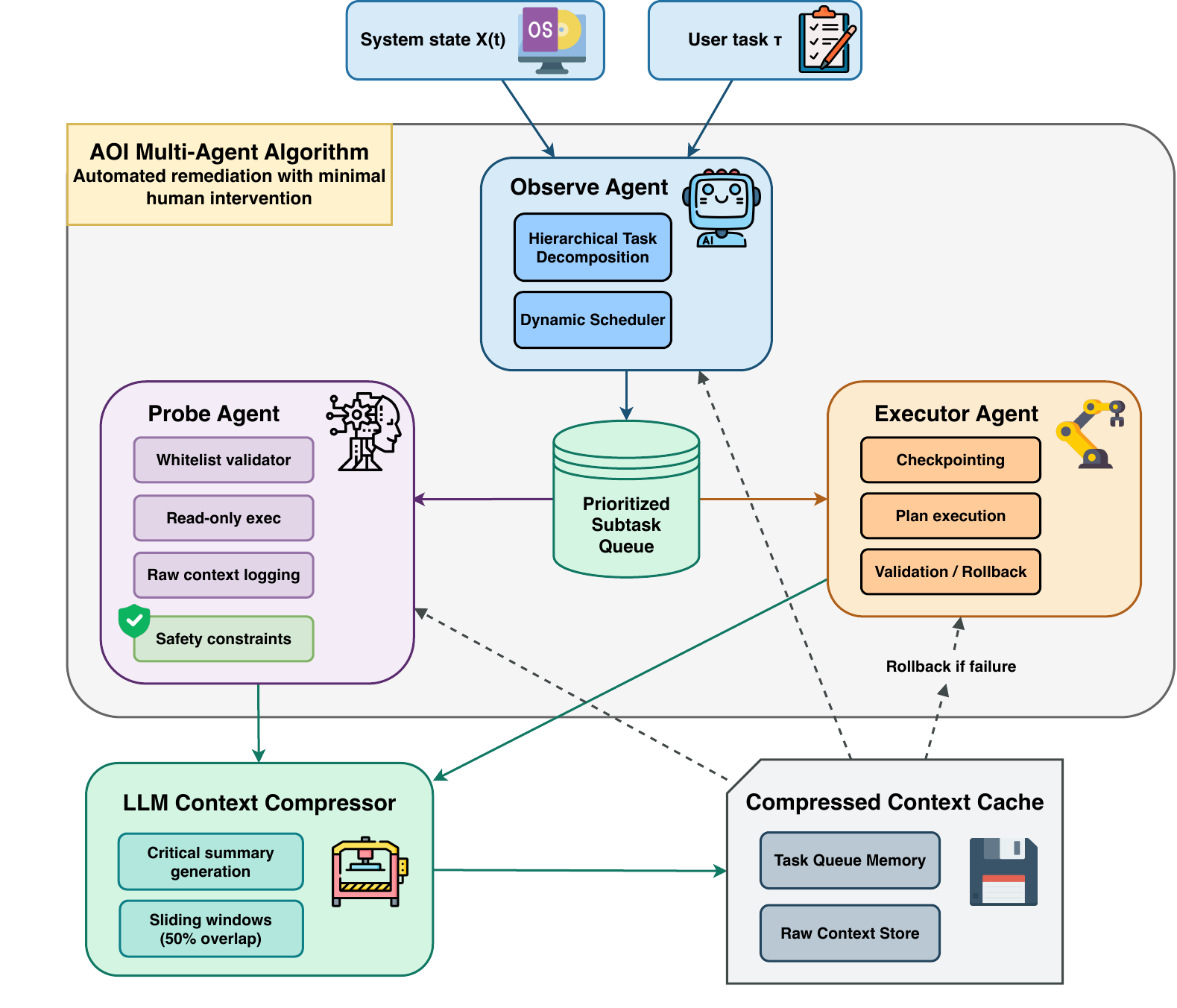}
\caption{AOI architecture. The Observer decomposes incidents and schedules work, the Probe performs safe read-only evidence gathering, and the Executor applies guarded recovery actions. A Context Compressor connects the agents to a three-layer memory system, allowing the framework to retain high-value operational evidence while reducing long-context overhead.}
\label{fig:aoi_detailed_architecture}
\end{figure*}

\begin{itemize}[left =0em]
\item \textbf{Multi-Agent Operational Architecture}: We propose a three-agent framework---Observer, Probe, and Executor---that separates coordination, read-only diagnosis, and guarded intervention.
\item \textbf{Memory-Aware Context Compression}: We introduce an LLM-based compressor with overlapping windows that reduces context size by over 70\% while preserving the information needed for diagnosis and recovery.
\item \textbf{Dynamic Task Scheduling}: We develop an adaptive scheduler that balances probing and execution using real-time state, task dependency, and historical incident signals.
\item \textbf{Three-Layer Memory Architecture}: We design hierarchical memory with raw context, task queues, and compressed context layers to support efficient retrieval and cross-incident reuse.
\end{itemize}

\begin{table*}[t!]
\centering
\caption{Comparison with representative automated operations methods. AOI is designed to combine the properties that are usually studied separately: multi-agent coordination, context compression, adaptive scheduling, LLM integration, explicit safety control, hierarchical memory, and formal analysis.}
\label{tab:related_work_comparison}
\small
\resizebox{\textwidth}{!}{
\renewcommand{\arraystretch}{0.95}
\begin{tabular}{|l|c|c|c|c|c|c|c|}
\hline
\textbf{Approach} & \textbf{Multi-Agent} & \textbf{Context} & \textbf{Dynamic} & \textbf{LLM} & \textbf{Safety} & \textbf{Memory} & \textbf{Theoretical} \\
 & \textbf{Coordination} & \textbf{Compression} & \textbf{Scheduling} & \textbf{Integration} & \textbf{Guarantees} & \textbf{Hierarchy} & \textbf{Foundation} \\
\hline
Expert Systems~\cite{xu2017belief,sarazin2021expert} & \xmark & \xmark & \xmark & \xmark & \checkmark & \xmark & \xmark \\
\hline
DeepLog~\cite{du2017deeplog} & \xmark & \xmark & \xmark & \xmark & \xmark & \xmark & Limited \\
\hline
LogAnomaly~\cite{meng2019loganomaly} & \xmark & \xmark & \xmark & \xmark & \xmark & \xmark & Limited \\
\hline
Traditional AIOps~\cite{dang2019aiops} & \xmark & Limited & \xmark & \xmark & Limited & \xmark & \xmark \\
\hline
MAS Automation~\cite{seitz2021automation} & \checkmark & \xmark & Limited & \xmark & Limited & \xmark & \xmark \\
\hline
Cloud-MAS~\cite{d2021designing} & \checkmark & Limited & Limited & \xmark & Limited & \xmark & \xmark \\
\hline
LLMLingua~\cite{jiang2023llmlingua} & \xmark & \checkmark & \xmark & \checkmark & \xmark & \xmark & Limited \\
\hline
HiAgent~\cite{hu2025hiagent} & Limited & \checkmark & \xmark & \checkmark & \xmark & \checkmark & \xmark \\
\hline
\textbf{AOI (Ours)} & \checkmark & \checkmark & \checkmark & \checkmark & \checkmark & \checkmark & \checkmark \\
\hline
\end{tabular}
}
\end{table*}

\section{Related Work}
\label{sec:related_work}
AOI sits at the intersection of automated IT operations, multi-agent coordination, tool-using language agents, and long-context memory. We review these areas with an emphasis on what they do and do not provide for incident response in cloud-native environments.

\subsection{Traditional Automated IT Operations}

Early efforts toward automating IT operations primarily relied on rule-based expert systems and deterministic diagnostic engines. Classical expert systems such as belief rule-based reasoning frameworks demonstrated effectiveness in structured industrial environments \cite{xu2017belief}. Similarly, knowledge-graph–driven expert systems have been applied to condition-based maintenance and operational optimization in industrial systems \cite{sarazin2021expert, muslewski2022expert}. These systems offered transparent and interpretable decision-making but required extensive manual rule engineering and were limited in their ability to adapt to evolving infrastructures.

Beyond rule-based expert systems, a line of research has investigated formal reliability and diagnosability properties of complex networked systems, providing theoretical foundations for fault detection and isolation. Prior studies have analyzed conditional diagnosability, connectivity, and fault tolerance of various interconnection networks under rigorous diagnostic models, demonstrating how structural properties directly impact system robustness and recovery guarantees. Such works offer valuable insights into how system topology and fault models influence safe operation and diagnosis, which motivates the safety-aware and reliability-driven design principles adopted in autonomous IT operations frameworks~\cite{wang2017g,wang2021connectivity,wang2024diagnosability,wang2017nature}.

With the rise of DevOps, automation workflows increasingly incorporated Infrastructure as Code and configuration-management tools. Systematic analyses of Infrastructure-as-Code research document progress in reproducible provisioning, deployment, and configuration control \cite{rahman2019systematic}. DevOps surveys likewise show that automation improves delivery velocity and operational consistency \cite{leite2019survey}. Yet these systems are primarily execution frameworks: they apply predefined procedures, but they do not decide which evidence to gather, when to stop probing, or whether a proposed repair is safe in a changing incident state. Recent work on infrastructure assistants highlights the same gap and calls for more reasoning-capable operational support \cite{yang2025cloudinfra}.

\subsection{Machine Learning in AIOps}

Machine learning–based approaches, collectively known as AIOps, have emerged to overcome the limitations of static rule-based systems. Surveys of AIOps methods describe a broad landscape of anomaly detection, diagnosis, event correlation, and automated response \cite{dang2019aiops,yeruva2023aiops,diaz2023joint,zhang2025survey}. Log-based anomaly detection has been especially influential. DeepLog \cite{du2017deeplog} and LogAnomaly \cite{meng2019loganomaly} learn behavioral patterns from logs and detect deviations at scale, offering a practical alternative to manual log inspection on datasets such as Loghub \cite{Loghub}.

Other AIOps work addresses predictive analytics, resource management, and explainable diagnosis \cite{zi2024time,chen2022explainable}. These models are valuable, but they often operate over a single modality or a fixed incident stage. In practice, an autonomous operations system must connect heterogeneous evidence, remember earlier observations, and decide whether the next action should collect more information or modify the system. AOI builds on AIOps prediction methods but treats them as part of a broader closed-loop recovery process.

\subsection{Multi-Agent Systems for Automated IT Operations}

Multi-agent systems (MAS) provide another promising paradigm for distributed automation. Foundational MAS literature highlights their advantages in decentralization, autonomy, and collaborative problem-solving \cite{wooldridge2009introduction}. In operational environments, MAS architectures have been applied to distributed monitoring and anomaly detection, allowing agents to independently observe and report on system states across heterogeneous infrastructures \cite{khosravifar2018anomaly}. Other works explore collaborative fault diagnosis using agent coordination strategies, enabling distributed decision-making and improved fault localization \cite{shao2021novel, xu2023graph}.

More recent research extends MAS to cloud-native and manufacturing environments. D'Aniello et al. demonstrate distributed operational management using cooperative agents in cloud-manufacturing scenarios \cite{d2021designing}, while adaptive resource-management studies show how MAS can respond to dynamic workloads \cite{seitz2021automation}. Xu et al. further illustrate MAS applications in autonomous supply chains, emphasizing task decomposition and coordination under uncertainty \cite{xu2024implementing}. However, existing MAS designs often have limited shared context, weak long-term memory, and little support for deciding when additional evidence is worth the delay. AOI targets these gaps through explicit role separation, shared compressed memory, and dynamic scheduling.

\subsection{LLM Agents and API Benchmarks}

Recent agent benchmarks have shifted evaluation from static question answering toward interaction with tools, APIs, software repositories, web environments, and operating systems. ReAct \cite{yao2023react}, Reflexion \cite{shinn2023reflexion}, and Tree of Thoughts \cite{yao2023tree} introduced influential patterns for reasoning-and-acting agents. API-centered benchmarks such as ToolLLM \cite{qin2023toolllm}, API-Bank \cite{li2023apibank}, Gorilla \cite{patil2023gorilla}, and StableToolBench \cite{guo2024stabletoolbench} evaluate tool selection and tool-use reliability. Broader agent benchmarks, including AgentBench \cite{liu2023agentbench}, WebArena \cite{zhou2023webarena}, OSWorld \cite{xie2024osworld}, SWE-bench \cite{jimenez2023swebench}, GAIA \cite{mialon2023gaia}, and AutoGen \cite{wu2023autogen}, make environment interaction and multi-agent communication measurable.

These benchmarks are important for general agent research, but they do not directly model operational risk in IT infrastructure. Their tasks are usually judged by functional success, whereas AIOps requires additional criteria such as MTTR, false-positive actions, rollback safety, information preservation, and context efficiency. AOI therefore uses agentic reasoning as a building block, but evaluates it in an operations-specific setting where acting too early can be unsafe and acting too late increases downtime.

\subsection{Context Management and Compression in Large Language Models}

Recent advances in large language models have opened new possibilities for operations automation, but they also make long-context management a central concern \cite{li2024surveyingmllm}. Long-document Transformer architectures such as Longformer \cite{beltagy2020longformer}, Reformer \cite{kitaev2020reformer}, and BigBird \cite{zaheer2020big} provide scalable attention mechanisms to extend context windows \cite{song2025transformersurvey}. However, they still face limitations when applied to continuously evolving, high-volume operational data streams.

To address context overload, several context compression and sparsification techniques have been proposed. LLMLingua \cite{jiang2023llmlingua} compresses long LLM contexts to accelerate inference while preserving essential semantics, whereas contextual sparsity approaches such as DejaVu prune redundant information during inference \cite{liu2023deja}. Complementary memory-augmented models, such as the Recurrent Memory Transformer (RMT) \cite{bulatov2022recurrent}, introduce persistent external memory to support long-range reasoning. Recent studies further reveal that LLMs exhibit position-dependent performance degradation in long contexts ("lost in the middle") \cite{liu2024lost,2024vilas}, and struggle with in-context exploration without explicit guidance \cite{krishnamurthy2024can,peng2024securingllm, 2025gittaskbench}.

While these methods improve efficiency and scalability, they are not designed specifically for IT operations, where the value of an event depends on incident stage, dependency structure, and the proposed repair action \cite{brewer2016sre}. Existing context mechanisms rarely prioritize information according to operational risk, and they usually do not connect compression quality to recovery metrics such as MTTR or false-positive actions \cite{lewis2020retrieval,borji2023categorical,liu2025comprehensive}. Emerging hierarchical memory architectures such as HiAgent \cite{hu2025hiagent} and retrieval-augmented in-context learning \cite{chen2024retrieval,yu2025spatial} offer promising directions, but their role in autonomous operations remains underexplored \cite{syed2024ai,diaz2023joint}.

\section{Theoretical Foundations}

We establish the theoretical foundations of the AOI framework through formal analysis of multi-agent coordination, dynamic scheduling convergence, and context compression stability.

\subsection{Markov Decision Process Formalization}

We model the AOI multi-agent coordination as a Decentralized Partially Observable Markov Decision Process (Dec-POMDP), defined by the tuple $\mathcal{M} = \langle \mathcal{N}, \mathcal{S}, \{\mathcal{A}_i\}, \mathcal{T}, \mathcal{R}, \{\Omega_i\}, \mathcal{O}, \gamma \rangle$, where:
\begin{itemize}
    \item $\mathcal{N} = \{Observer, Probe, Executor\}$ is the set of agents
    \item $\mathcal{S}$ is the joint state space of the IT infrastructure
    \item $\mathcal{A}_i$ is the action space for agent $i$
    \item $\mathcal{T}: \mathcal{S} \times \mathcal{A} \times \mathcal{S} \rightarrow [0,1]$ is the state transition function
    \item $\mathcal{R}: \mathcal{S} \times \mathcal{A} \rightarrow \mathbb{R}$ is the shared reward function
    \item $\Omega_i$ is the observation space for agent $i$
    \item $\mathcal{O}: \mathcal{S} \times \mathcal{A} \times \Omega \rightarrow [0,1]$ is the observation function
    \item $\gamma \in [0,1)$ is the discount factor
\end{itemize}

The optimal joint policy $\pi^* = \{\pi_i^*\}_{i \in \mathcal{N}}$ maximizes the expected cumulative discounted reward:
\begin{equation}
\pi^* = \arg\max_{\pi} \mathbb{E}\left[\sum_{t=0}^{\infty} \gamma^t \mathcal{R}(s_t, a_t) \mid \pi \right]
\end{equation}

\subsection{Dynamic Scheduling Convergence Analysis}

We analyze the convergence of the dynamic scheduling strategy through a potential-based finite-time argument. Let the scheduling state at round $t$ be
\begin{equation}
\sigma_t = (q_t, c_t, h_t),
\end{equation}
where $q_t$ is the task queue state, $c_t$ is the compressed context, and $h_t$ is the historical pattern embedding. Let $\mathcal{A}$ denote the finite action set of feasible scheduling decisions.

We impose the following assumptions. The action set $\mathcal{A}$ is finite. The one-step reward function $r(\sigma,a)$ is bounded in $[0,1]$ and is $L$-Lipschitz in $\sigma$. The potential function
\begin{equation}
\Phi(\sigma_t) = \sum_{\tau \in q_t} w(\tau) d(\tau,t)
\end{equation}
is nonnegative and lower bounded by $0$. The scheduling update satisfies the stochastic drift inequality
\begin{equation}
\begin{aligned}
\mathbb{E}[\Phi(\sigma_{t+1}) - \Phi(\sigma_t) \mid \mathcal{F}_t]
&\leq -\eta \|\nabla \Phi(\sigma_t)\|^2 \\
&\quad + \frac{\eta^2 G^2}{2} + \beta_t,
\end{aligned}
\end{equation}
where $\mathcal{F}_t$ is the natural filtration up to round $t$, $\eta > 0$ is the step size, $G$ is a uniform bound on the stochastic gradient norm, and $\beta_t$ is the exploration penalty. We further assume that the exploration bonus is UCB-type, so that with probability at least $1-\delta$, the cumulative regret after $T$ rounds satisfies
\begin{equation}
\begin{aligned}
R_T &= \sum_{t=1}^T \bigl(r(\sigma_t,a_t^\star) - r(\sigma_t,a_t)\bigr)\\ &\leq C \sqrt{|\mathcal{A}| T \log(|\mathcal{A}|/\delta)},
\end{aligned}
\end{equation}
where $a_t^\star$ denotes the best static action in hindsight. Finally, assume that the expected task completion time $J(\pi)$ is controlled by the time-averaged potential, in the sense that there exist constants $m,M > 0$ and $b \geq 0$ such that
\begin{equation}
\begin{aligned}
m \frac{1}{T} \sum_{t=1}^T \mathbb{E}[\Phi(\sigma_t)] - b
&\leq J(\pi) \\
&\leq M \frac{1}{T} \sum_{t=1}^T \mathbb{E}[\Phi(\sigma_t)] + b.
\end{aligned}
\end{equation}

Under these assumptions, we obtain the following result.

\begin{theorem}[Finite-time convergence of dynamic scheduling]
\label{thm:scheduling_convergence}
Let $\pi_T$ be the AOI dynamic scheduling policy run for $T$ rounds with exploration parameter $\lambda \in (0,1)$. Suppose that $\beta_t \leq c_0 \lambda / \sqrt{t}$ for some constant $c_0 > 0$. Then, with probability at least $1-\delta$,
\begin{equation}
\begin{aligned}
&J(\pi_T) - J(\pi^\star_{\mathrm{static}})\\
\leq {}& C_1 \frac{\Phi(\sigma_1)}{T} + C_2 \eta G^2 + C_3 \lambda \\
&+ C_4 \sqrt{\frac{|\mathcal{A}| \log(|\mathcal{A}|/\delta)}{T}},
\end{aligned}
\end{equation}
where $\pi^\star_{\mathrm{static}}$ denotes the optimal static scheduling policy and $C_1,C_2,C_3,C_4 > 0$ are problem-dependent constants. In particular, for any $\epsilon > 0$, if $\eta$ and $\lambda$ are chosen so that
\begin{equation}
C_2 \eta G^2 + C_3 \lambda \leq \frac{\epsilon}{2},
\end{equation}
and
\begin{equation}
T \geq \frac{4 C_4^2 |\mathcal{A}| \log(|\mathcal{A}|/\delta)}{\epsilon^2},
\end{equation}
then
\begin{equation}
J(\pi_T) \leq J(\pi^\star_{\mathrm{static}}) + \epsilon
\end{equation}
with probability at least $1-\delta$.
\end{theorem}

Theorem~\ref{thm:scheduling_convergence} provides a finite-time guarantee for the proposed dynamic scheduling strategy. It shows that the expected completion-time gap between AOI and the best static scheduler decreases with the number of scheduling rounds, up to controllable optimization and exploration terms. This result supports our design choice of using adaptive scheduling rather than fixed probing or execution rules: as more operational feedback is collected, the scheduler can approach the performance of the best static policy while still retaining exploration for uncertain system states. The proof is provided in Section~\ref{sec:proof_scheduling_convergence}.

\subsection{Context Compression Stability}

We analyze the stability of the LLM-based context compression through a Lyapunov argument. Let $\mathcal{C}_t$ denote the context state at time $t$, and let $\mathcal{F}: \mathcal{C} \to \mathcal{C}$ denote the compression operator induced by the sliding-window summarization mechanism.

We assume that there exists an equilibrium context state $\mathcal{C}^*$ such that
\begin{equation}
\mathcal{F}(\mathcal{C}^*) = \mathcal{C}^*.
\end{equation}
We further assume that, in the RKHS norm induced by the LLM embedding, the compression operator satisfies
\begin{equation}
\|\mathcal{F}(\mathcal{C}) - \mathcal{C}^*\|_{\mathcal{H}} \leq \alpha \|\mathcal{C} - \mathcal{C}^*\|_{\mathcal{H}} + \epsilon_{\mathrm{LLM}},
\end{equation}
where $\alpha \in (0,1)$ is an effective contraction factor determined by the overlap ratio $\rho$, and $\epsilon_{\mathrm{LLM}} \geq 0$ is the approximation error of the LLM compressor. In particular, larger overlap corresponds to smaller $\alpha$, and we assume $\rho > 0.5$ so that $\alpha < 1$.

\begin{theorem}[Practical stability of context compression]
\label{thm:compression_stability}
Suppose the compression operator satisfies
\begin{equation}
\|\mathcal{F}(\mathcal{C}) - \mathcal{C}^*\|_{\mathcal{H}} \leq \alpha \|\mathcal{C} - \mathcal{C}^*\|_{\mathcal{H}} + \epsilon_{\mathrm{LLM}}
\end{equation}
for some $\alpha \in (0,1)$. Then the context compression mechanism is practically stable. More precisely, there exists a Lyapunov function $V: \mathcal{C} \to \mathbb{R}_{\geq 0}$ such that
\begin{equation}
\begin{aligned}
V(\mathcal{F}(\mathcal{C})) - V(\mathcal{C}) &\leq -\frac{1-\alpha^2}{2}\|\mathcal{C} - \mathcal{C}^*\|_{\mathcal{H}}^2 \\&+ \frac{\alpha \epsilon_{\mathrm{LLM}}}{1-\alpha}\|\mathcal{C} - \mathcal{C}^*\|_{\mathcal{H}} + \frac{1}{2}\epsilon_{\mathrm{LLM}}^2.
\end{aligned}
\end{equation}
Consequently, the compressed context iterates remain ultimately bounded, and
\begin{equation}
\limsup_{t \to \infty}\|\mathcal{C}_t - \mathcal{C}^*\|_{\mathcal{H}} \leq \frac{\epsilon_{\mathrm{LLM}}}{1-\alpha}.
\end{equation}
In particular, if $\epsilon_{\mathrm{LLM}} = 0$, then the equilibrium $\mathcal{C}^*$ is asymptotically stable.
\end{theorem}

Theorem~\ref{thm:compression_stability} justifies the stability of the LLM-based context compression module. It shows that if the compression operator is contractive up to bounded LLM approximation error, then the compressed context remains ultimately bounded around an equilibrium context state. This supports the use of repeated sliding-window compression in long-running operations: compression reduces context size without causing unbounded semantic drift, and the retained context error is controlled by $\epsilon_{\mathrm{LLM}}/(1-\alpha)$. The proof is provided in Section~\ref{sec:proof_compression_stability}.

\subsection{Information Preservation Bound}

We further analyze whether the context compressor preserves task-relevant operational information.
Let $\mathcal{C}$ denote the raw operational context, $\mathcal{Y}$ denote the downstream operational outcome, such as the correct diagnosis or recovery action, and let
\begin{equation}
\mathcal{C}_{\mathrm{comp}} = \mathcal{F}_{w,\rho}(\mathcal{C})
\end{equation}
be the compressed context produced by the sliding-window compressor with window size $w$ and overlap ratio $\rho$.
Since $\mathcal{C}_{\mathrm{comp}}$ is a deterministic or randomized summary of $\mathcal{C}$, the data processing inequality implies that compression cannot increase the information about $\mathcal{Y}$.
The key question is therefore how much information is lost.

\begin{theorem}[Task-Relevant Information Preservation]
\label{thm:info_preservation}
Let $\mathcal{C}$ be the raw operational context, $\mathcal{C}_{\mathrm{comp}}=\mathcal{F}_{w,\rho}(\mathcal{C})$ be the compressed context, and $\mathcal{Y}$ be the operational outcome. Suppose the compression operator satisfies the task-relevance loss condition
\begin{equation}
I(\mathcal{Y};\mathcal{C}\mid \mathcal{C}_{\mathrm{comp}})
\leq
\varepsilon_{w,\rho},
\label{eq:task_relevance_loss}
\end{equation}
where
\begin{equation}
\varepsilon_{w,\rho}
\leq
\frac{K}{w\rho}
\label{eq:eps_window_overlap}
\end{equation}
for some constant $K>0$. Then the compressed context preserves the mutual information between the raw context and the operational outcome up to an additive error:
\begin{equation}
I(\mathcal{C}_{\mathrm{comp}};\mathcal{Y})
\geq
I(\mathcal{C};\mathcal{Y}) - \varepsilon_{w,\rho}.
\label{eq:additive_info_bound}
\end{equation}
Equivalently,
\begin{equation}
I(\mathcal{C}_{\mathrm{comp}};\mathcal{Y})
\geq
I(\mathcal{C};\mathcal{Y}) - \frac{K}{w\rho}.
\label{eq:window_overlap_bound}
\end{equation}
Moreover, if $I(\mathcal{C};\mathcal{Y})>0$, then
\begin{equation}
I(\mathcal{C}_{\mathrm{comp}};\mathcal{Y})
\geq
\left(1-\frac{K}{w\rho I(\mathcal{C};\mathcal{Y})}\right)
I(\mathcal{C};\mathcal{Y}).
\label{eq:relative_info_bound}
\end{equation}
\end{theorem}

The proof is provided in Section~\ref{thm:proof3}. Theorem~\ref{thm:info_preservation} shows that AOI's context compressor can reduce context length while preserving task-relevant operational information. The bound depends on the conditional information loss $I(\mathcal{Y};\mathcal{C}\mid \mathcal{C}_{\mathrm{comp}})$, which measures the part of the raw context that remains useful for predicting the operational outcome but is not retained in the compressed context. The sliding-window design improves this bound through two mechanisms: larger windows include more local diagnostic evidence, while overlap reduces boundary loss when fault signatures span adjacent windows. Therefore, increasing the window size $w$ or overlap ratio $\rho$ decreases the upper bound on information loss. This theoretical result explains our empirical observation in Table~\ref{tab:param_sensitivity_window}, where larger window sizes improve information preservation while still enabling strong context compression.
\section{Methodology}\label{sec:methodology}
\subsection{Problem Statement and Framework Overview}
AOI is designed for incident-response settings in which an operator-level goal must be achieved under incomplete and changing system state. Unlike a static anomaly detector, the framework must repeatedly decide whether to observe, probe, execute, or wait. Four design challenges follow naturally:
\begin{itemize}[left = 0em]
    \item \textbf{Information overload}: compress high-volume operational context while preserving evidence needed for diagnosis and remediation.
    \item \textbf{Task coordination}: schedule probing and execution so that information gain, downtime reduction, and operational risk are jointly considered.
    \item \textbf{Context retention}: store and retrieve incident context across long-running operations and recurring failure patterns.
    \item \textbf{Operational safety}: guard state-changing actions with validation and rollback so that remediation does not introduce avoidable failures.
\end{itemize}

AOI addresses these challenges with four coupled components. The Observer decomposes incident goals and maintains the global plan. The Probe gathers read-only evidence under least-privilege constraints. The Executor applies validated recovery actions with checkpointing and rollback. The Context Compressor converts raw logs, metrics, and execution traces into compact memory states that remain useful for later decisions. Figure~\ref{fig:aoi_detailed_architecture} shows how these modules interact through the three-layer memory system.
\subsection{Problem Formulation}
Let $\mathcal{S} = \{s_1, s_2, ..., s_n\}$ represent the set of system components under management, where each component $s_i$ has a state vector $\mathbf{x}_i(t) \in \mathbb{R}^d$ at time $t$. Given a user task $\tau$ and current system state $\mathbf{X}(t) = [\mathbf{x}_1(t), \mathbf{x}_2(t), ..., \mathbf{x}_n(t)]^T$, our objective is to find an optimal sequence of actions $\mathcal{A} = \{a_1, a_2, ..., a_k\}$ that transforms the system to a desired target state $\mathbf{X}^*$ while minimizing the operational cost function:$\mathcal{J}(\mathcal{A}) = \alpha \cdot T_{completion} + \beta \cdot C_{resource} + \gamma \cdot R_{risk}$, where $T_{completion}$ is the Mean Time to Resolution (MTTR), $C_{resource}$ represents computational and network overhead, $R_{risk}$ quantifies the operational risk score of potential system instability or SLA violations, and $\alpha$, $\beta$, $\gamma$ are weighting parameters.
\subsection{AOI Framework Architecture}
Figure~\ref{fig:aoi_detailed_architecture} shows the AOI architecture, which contains four core modules and a three-layer memory system.
\subsubsection{Observer Agent}
The Observer Agent serves as the central coordination unit responsible for task analysis, decomposition, and strategic decision-making. Its core functionalities include:
\textbf{Task Decomposition Algorithm}: Given a complex task $\tau$, the Observer employs a hierarchical decomposition strategy shown in Algorithm~\ref{alg:task_decomposition}.

\begin{algorithm}[htbp]
\caption{Dynamic Task Decomposition}
\label{alg:task_decomposition}
\begin{algorithmic}[1]
\REQUIRE Task $\tau$, System State $\mathbf{X}(t)$, Context $C_{compressed}$
\ENSURE Subtask Queue $\mathcal{Q}_{subtasks}$
\STATE Initialize $\mathcal{Q}_{subtasks} \leftarrow \emptyset$
\STATE Analyze task complexity $\mathcal{C}(\tau)$
\IF{$\mathcal{C}(\tau) > \theta_{complex}$}
    \STATE Decompose $\tau$ into atomic subtasks $\{\tau_1, \tau_2, ..., \tau_m\}$
    \FOR{each $\tau_i$ in decomposed tasks}
        \STATE Determine task type: $type(\tau_i) \in \{probe, execute\}$
        \STATE Estimate resource requirements $R(\tau_i)$
        \STATE Assign priority $P(\tau_i)$ based on dependency analysis
        \STATE Enqueue $(\tau_i, type(\tau_i), R(\tau_i), P(\tau_i))$ to $\mathcal{Q}_{subtasks}$
    \ENDFOR
\ELSE
    \STATE Enqueue $\tau$ directly to $\mathcal{Q}_{subtasks}$
\ENDIF
\RETURN $\mathcal{Q}_{subtasks}$
\end{algorithmic}
\end{algorithm}

\textbf{Dynamic Scheduling Strategy}: The Observer implements an adaptive scheduling algorithm that balances exploration (probing) and exploitation (execution) based on current information sufficiency:
\begin{equation}
\resizebox{\columnwidth}{!}{$
S(t)=\arg\max_{a\in\{\mathrm{probe}, \mathrm{execute}\}}\mathbb{E}[\mathrm{Reward}(a, \mathbf{X}(t), C(t))]
$}
\end{equation}
\noindent where the expected reward combines the value of additional evidence with the expected progress toward the target state. A \texttt{probe} action is rewarded when it reduces system-state uncertainty, while an \texttt{execute} action is rewarded when it advances the system toward $\mathbf{X}^*$ under acceptable risk.
\subsubsection{Probe Agent}
The Probe Agent is a strictly read-only component for safe system information gathering. 
\textbf{Safety-by-Design}: It follows the \textbf{least-privilege} principle to prevent unintended modifications during diagnostics. 
\textbf{Safety Constraints}: The agent is limited to non-destructive query and read operations (e.g., database introspection and log/file reads), while any state-changing actions (e.g., writes, deletions, schema changes, or destructive commands) are explicitly prohibited. 
File access remains read-only, and network interactions are restricted to query-type requests.

\textbf{Information Collection Strategy}: The Probe employs a structured probing strategy (Algorithm~\ref{alg:safe_probing}).

\begin{algorithm}[htbp]
\caption{Safe Information Probing}
\label{alg:safe_probing}
\begin{algorithmic}[1]
\REQUIRE Probe Task $\tau_{probe}$, Target System $\mathcal{S}$
\ENSURE Raw Information $I_{raw}$
\STATE Initialize $I_{raw} \leftarrow \emptyset$
\STATE Generate probe script $script \leftarrow GenerateScript(\tau_{probe})$
\STATE Validate script safety: $ValidateSafety(script)$
\IF{script is safe}
    \FOR{each command $cmd$ in script}
        \STATE Execute $result \leftarrow ExecuteReadOnly(cmd, \mathcal{S})$
        \STATE Append $result$ to $I_{raw}$
        \IF{error occurred}
            \STATE Log error and continue with next command
        \ENDIF
    \ENDFOR
\ELSE
    \STATE Reject unsafe script and request revision
\ENDIF
\STATE Store $I_{raw}$ in Memory-Raw Context
\RETURN $I_{raw}$
\end{algorithmic}
\end{algorithm}
\subsubsection{Executor Agent}
The Executor Agent implements system modifications to achieve target operational outcomes. Unlike the Probe Agent, it can modify system state under coordination with the Observer.
\textbf{Conservative Execution Strategy}: The Executor applies changes only after validation and checkpoint creation. If a high-severity failure is detected after execution, the system rolls back to the most recent checkpoint (Algorithm~\ref{alg:safe_execution}).
\begin{algorithm}[htbp]
\caption{Safe System Execution}
\label{alg:safe_execution}
\begin{algorithmic}[1]
\REQUIRE Execution Task $\tau_{exec}$, Target System $\mathcal{S}$
\ENSURE Execution Result $R_{exec}$
\STATE Create checkpoint $checkpoint \leftarrow CreateCheckpoint(\mathcal{S})$
\STATE Generate execution plan $plan \leftarrow GeneratePlan(\tau_{exec})$ \COMMENT{Plan generation includes pre-execution validation and dependency checks}
\FOR{each action $a$ in plan}
    \IF{RequiresSystemInfo(a)}
        \STATE Query Probe Agent for current state
        \STATE Update action parameters based on current state
    \ENDIF
    \STATE Execute $result \leftarrow ExecuteAction(a, \mathcal{S})$
    \STATE Validate execution result
    \IF{critical failure detected}
        \STATE Initiate rollback to $checkpoint$
        \STATE BREAK
    \ENDIF
\ENDFOR
\STATE Store execution results in Memory-Raw Context
\RETURN $R_{exec}$
\end{algorithmic}
\end{algorithm}
\subsubsection{Context Compressor}
The Context Compressor uses large language models to process and compress operational context while preserving diagnostic information.
\textbf{Semantic-Aware Compression}: Unlike keyword-based or statistical compression methods, the compressor identifies error patterns, resource thresholds, dependency evidence, and causal relations while discarding redundant or low-signal data.
\textbf{Sliding Window Compression}: We implement a sliding window mechanism with 50\% overlap to preserve context continuity when fault evidence spans two adjacent windows: $
\mathrm{Window_i} = [start_i, start_i + window_{size}]$
where $start_i = i \times (window_{size} \times overlap_{ratio})$ and $overlap_{ratio} = 0.5$.
\textbf{Intelligent Compression Strategy}: Algorithm~\ref{alg:context_compression} prioritizes information based on criticality and merges overlapping summaries before the compressed context is written back to memory.

\begin{algorithm}[htbp]
\caption{LLM-based Context Compression}
\label{alg:context_compression}
\begin{algorithmic}[1]
\REQUIRE Raw Context $C_{raw}$, Compression Target Ratio $r_{target}$
\ENSURE Compressed Context $C_{compressed}$
\STATE Split $C_{raw}$ into sliding windows $\{W_1, W_2, ..., W_k\}$
\STATE Initialize $C_{compressed} \leftarrow \emptyset$
\FOR{each window $W_i$}
    \STATE Identify critical information: $I_{critical} \leftarrow \mathrm{ExtractCritical}(W_i)$ \COMMENT{Using LLM to extract fault signatures, error codes, and performance anomalies}
    \STATE Generate summary: $S_i \leftarrow \mathrm{LLMSummarize}(W_i, I_{critical})$
    \STATE Append $S_i$ to $C_{compressed}$
\ENDFOR
\STATE Merge overlapping summaries: $C_{merged} \leftarrow \mathrm{MergeOverlaps}(C_{compressed})$
\STATE Apply secondary compression if needed to meet $r_{target}$
\RETURN $C_{merged}$
\end{algorithmic}
\end{algorithm}

\subsection{Three-Layer Memory Architecture}
Our memory management system consists of three specialized layers designed following the principle of \textbf{data lifecycle management}, where information flows from high-volume, short-term raw storage to condensed, long-term strategic memory, optimizing both retrieval speed and storage efficiency.

\textbf{Layer 1: Raw Context Storage --- }Stores unprocessed results from Probe and Executor agents with 24-hour retention. \textbf{(Feeds into Context Compressor)}.
\textbf{Layer 2: Task Queue Management --- } Maintains structured execution instructions with lifecycle management. \textbf{(Managed by Observer and executed by Executor)}.
\textbf{Layer 3: Compressed Context Cache --- } Houses LLM-processed information with 7-day retention for strategic decision-making. \textbf{(Populated by Context Compressor and queried by Observer)}.
\subsection{Complexity Analysis}
The computational complexity of the AOI framework is primarily determined by the LLM-based context compression module, which exhibits a time complexity of $O(n \cdot w)$, where $n$ denotes the context length and $w$ represents the compression window size. The space complexity is $O(k \cdot w)$, with $k$ corresponding to the number of overlapping sliding windows. Benefiting from the distributed multi-agent design, the framework achieves near-linear scalability with respect to the number of concurrent tasks, as each agent operates independently with minimal inter-agent communication overhead.
\section{Experiments}\label{sec:experiments}
\subsection{Experimental Setup}
All experiments were conducted in a controlled cloud environment that emulates large-scale IT operations while allowing reproducible fault injection and recovery evaluation. 
The experimental cluster consisted of eight AWS EC2 \texttt{c5.4xlarge} instances, each equipped with 16 vCPUs and 32~GB of RAM, supported by high-IOPS persistent storage. 
All nodes were interconnected through a 10~Gbps network with adjustable latency injection to simulate diverse communication conditions. 
The software stack was built on Ubuntu~20.04~LTS, running Docker and Kubernetes for container orchestration. 
System monitoring and telemetry were implemented via an integrated observability framework composed of Prometheus, Grafana, and the ELK stack~\cite{Prometheus,Grafana,ELK}. 
The large language model backend uses GPT-4 for operations-oriented context understanding and summarization~\cite{GPT4}. 
This setting lets us evaluate AOI under the same telemetry, resource limits, and fault-injection schedule as the baselines, so performance differences reflect the framework design rather than differences in infrastructure access.
\subsection{Datasets and Scenarios}
We evaluate AOI on complementary simulated and real-world operational settings.
The \textbf{AIOpsLab Simulation Environment} contains 1,000 fault-injection scenarios covering 50 fault types, ranging from isolated failures to cascading faults, each annotated with ground-truth resolutions and post-recovery states~\cite{AIOpsLab}.
To assess practical applicability beyond simulation, we additionally use \textbf{Loghub}, which aggregates production logs from distributed storage systems, supercomputing systems, OpenStack deployments, and other enterprise infrastructures~\cite{Loghub}.
Logs are preprocessed via anonymization, parsing, and scenario segmentation to align with AOI’s task structure.
Across both datasets, evaluation focuses on four representative operational event categories: service failure recovery, performance degradation, configuration drift, and security incidents.
\subsection{Baseline Methods}
We compare AOI with four reproducible baselines that cover the main design choices in automated operations. The \textbf{Rule-based Expert System} (RES) follows deterministic runbooks and expert rules, representing classical knowledge-engineered automation \cite{xu2017belief,sarazin2021expert,rahman2019systematic}. The \textbf{Traditional AIOps Pipeline} (TAP) combines anomaly detection with scripted remediation, following the dominant detect-then-act structure in AIOps \cite{dang2019aiops,du2017deeplog,meng2019loganomaly,Loghub}. The \textbf{Single-Agent LLM} (SA-LLM) baseline gives one language-model agent access to the same observations and tools, reflecting tool-use settings studied in ReAct, ToolLLM, API-Bank, Gorilla, and StableToolBench \cite{yao2023react,qin2023toolllm,li2023apibank,patil2023gorilla,guo2024stabletoolbench}. The \textbf{Baseline Multi-Agent System} (B-MAS) uses multiple cooperating agents but removes AOI's context compression, three-layer memory, and adaptive scheduling, following standard multi-agent coordination and conversation frameworks \cite{wooldridge2009introduction,d2021designing,wu2023autogen}. Together, these baselines separate the value of automation rules, predictive AIOps, LLM tool use, multi-agent coordination, and AOI's proposed memory-scheduling design.
\subsection{Evaluation Metrics}
We evaluate both recovery quality and operational safety. The primary metrics are Task Success Rate (TSR), Mean Time to Resolution (MTTR), Context Compression Ratio (CCR), and Information Preservation Score (IPS). TSR measures whether the incident is resolved; MTTR measures recovery speed; CCR captures how much operational context is removed; and IPS measures whether the compressed context retains diagnosis- and recovery-relevant evidence. Secondary metrics include False Positive Rate (FPR), Resource Utilization Efficiency (RUE), Scalability Index (SI), and System Safety Score (SSS). These metrics are chosen because autonomous operations should not be judged only by correctness: a useful system must resolve incidents quickly, avoid harmful interventions, preserve the evidence needed for later decisions, and remain stable under concurrent load.
\subsection{Implementation Details}
Each agent is implemented as a containerized microservice in Python~3.9 using FastAPI for inter-agent communication~\cite{FastAPI}. 
Context compression integrates GPT-4 with structured operational instructions, incurring an average latency of 2.3 seconds per compression~\cite{GPT4}. 
Memory management relies on a Redis Cluster for distributed storage, failover, and persistence. 
Operational safety is ensured via formal TLA+ specifications and runtime monitoring with custom safety rules~\cite{TLAPLUS}.
\section{Results}\label{sec:results}
\subsection{Main Experimental Results}
Table~\ref{tab:main_results} shows that AOI improves both incident resolution and operational safety. AOI reaches a 94.2\% task success rate, exceeding the strongest baseline by 7.8 percentage points and the rule-based baseline by 26.4 percentage points. It also reduces MTTR to 22.1 minutes, a 34.4\% reduction relative to B-MAS. The context metrics explain why the improvement is not only a byproduct of using a stronger model: AOI removes 72.4\% of the raw context while preserving 92.8\% of critical information, allowing the agents to keep high-signal evidence available across the recovery loop. The same design also reduces false-positive actions to 3.1\% and raises the system safety score to 96.7\%, which is essential for autonomous operation in production-like settings.
\begin{table*}[t!]
\centering
\caption{Main results on the AIOpsLab simulation. Higher is better for TSR, CCR, IPS, RUE, SI, and SSS; lower is better for MTTR and FPR. AOI achieves the best recovery quality and safety while retaining high-value context under strong compression.}
\label{tab:main_results}
\resizebox{\textwidth}{!}{
\begin{tabular}{|l|c|c|c|c|c|c|c|c|}
\hline
\textbf{Method} & \textbf{TSR (\%)} & \textbf{MTTR (min)} & \textbf{CCR (\%)} & \textbf{IPS (\%)} & \textbf{FPR (\%)} & \textbf{RUE} & \textbf{SI} & \textbf{SSS (\%)} \\
\hline
Rule-based Expert System (RES) & 67.8 & 54.3 & -- & -- & 9.2 & 0.71 & 0.62 & 78.5 \\
\hline
Traditional AIOps Pipeline (TAP) & 75.1 & 42.6 & 35.7 & 81.4 & 7.8 & 0.76 & 0.74 & 82.3 \\
\hline
Single-Agent LLM (SA-LLM) & 81.5 & 38.2 & 52.9 & 86.7 & 6.9 & 0.79 & 0.77 & 88.6 \\
\hline
Baseline Multi-Agent System (B-MAS) & 86.4 & 33.7 & 61.8 & 89.3 & 5.6 & 0.82 & 0.81 & 90.1 \\
\hline
\textbf{Our AOI Framework} & \textbf{94.2} & \textbf{22.1} & \textbf{72.4} & \textbf{92.8} & \textbf{3.1} & \textbf{0.89} & \textbf{0.93} & \textbf{96.7} \\
\hline
\end{tabular}
}
\end{table*}

\begin{figure}[!ht]
\centering
\includegraphics[width=\columnwidth]{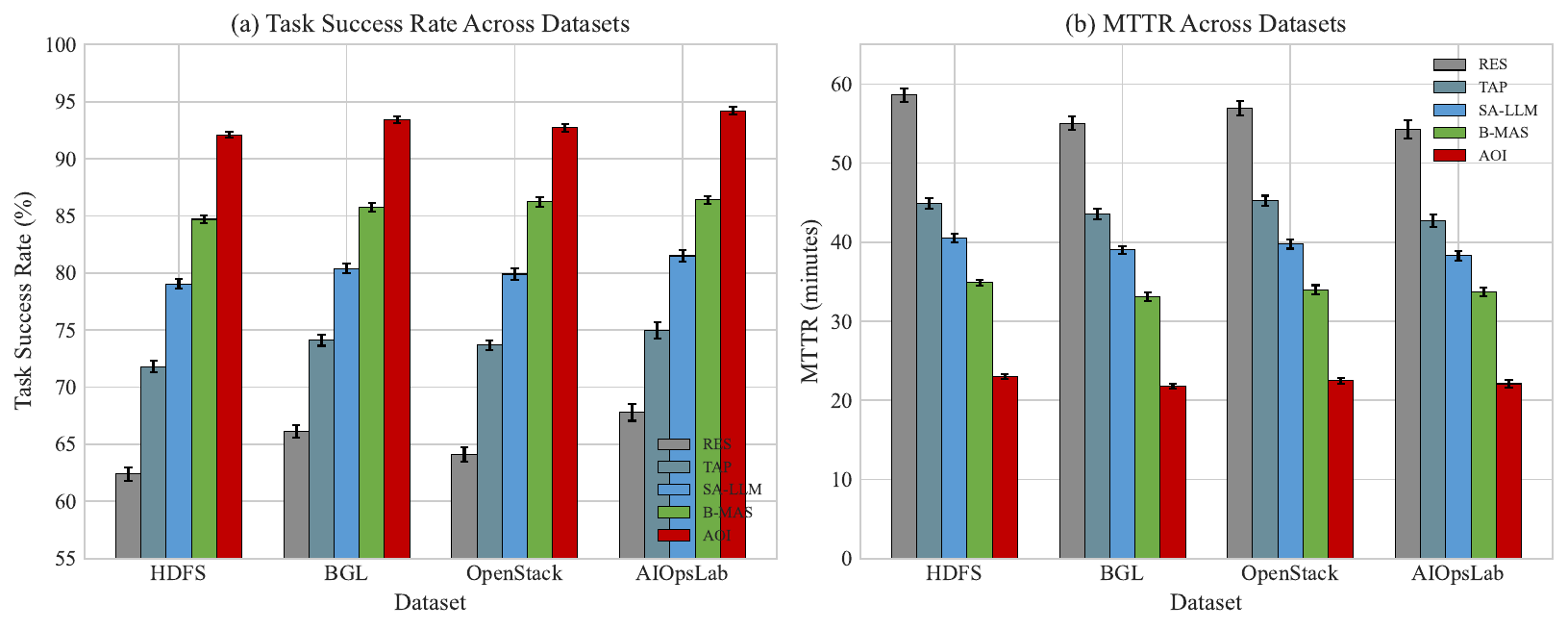}
\caption{Cross-dataset performance across HDFS, BGL, OpenStack, and AIOpsLab. AOI remains consistently above the baselines across real-world log datasets and simulated operational incidents, indicating that the gains are not tied to a single telemetry distribution.}
\label{fig:cross_dataset}
\end{figure}

We evaluate cross-dataset generalization on four heterogeneous benchmarks—HDFS, BGL, OpenStack, and AIOpsLab—covering both real-world and simulated operational environments.
As shown in Figure~\ref{fig:cross_dataset}, AOI consistently outperforms all baselines across datasets, indicating that the method transfers across different telemetry sources.
Quantitative results in Table~\ref{tab:cross_dataset_summary} further confirm this trend: AOI achieves the highest Task Success Rate (TSR) on all datasets (92.1--94.2\%) while reducing Mean Time To Resolution (MTTR) to around 22 minutes, clearly improving over the strongest baseline (B-MAS).
Notably, the performance gains are stable across diverse log characteristics and system settings, demonstrating that AOI generalizes well across operational domains rather than overfitting to a specific dataset.

\subsection{Ablation Studies}
Table~\ref{tab:ablation_results} isolates the contribution of AOI's main components, and Table~\ref{tab:ablation_detailed} reports the extended safety ablation. Removing the context compressor lowers TSR by 5.7 percentage points and increases MTTR by 7.7 minutes, showing that compression helps the agents focus on diagnostic evidence rather than telemetry noise. Removing dynamic scheduling produces the largest MTTR increase among single-module ablations, confirming that AOI benefits from deciding when to probe and when to execute. Replacing the three-layer memory reduces both task success and compression effectiveness, indicating that memory structure matters for reusing context across an incident. The single-agent variant is lower on all three metrics, supporting the choice to separate observation, probing, and execution.
\begin{table}[!ht]
\centering
\caption{Component ablation on AIOpsLab. Each variant removes one design element from AOI. The full framework provides the strongest combination of task success, fast recovery, and context compression.}
\label{tab:ablation_results}
\resizebox{\columnwidth}{!}{
\begin{tabular}{|l|c|c|c|}
\hline
\textbf{Configuration} & \textbf{TSR (\%)} & \textbf{MTTR (min)} & \textbf{CCR (\%)} \\
\hline
AOI (Full) & \textbf{94.2} & \textbf{22.1} & \textbf{72.4} \\
\hline
w/o Context Compressor & 88.5 & 29.8 & N/A \\
\hline
w/o Dynamic Scheduling & 90.1 & 26.7 & 70.5 \\
\hline
w/o Three-Layer Memory & 89.4 & 27.5 & 65.2 \\
\hline
Single-Agent Version & 84.6 & 31.4 & 60.3 \\
\hline
\end{tabular}
}
\end{table}
\subsection{Parameter Sensitivity Analysis}
We analyze the sensitivity of AOI to three key parameters: context compression window size, scheduling balance factor, and memory retention policy.

\begin{figure}[!ht]
\centering
\includegraphics[width=0.8\columnwidth]{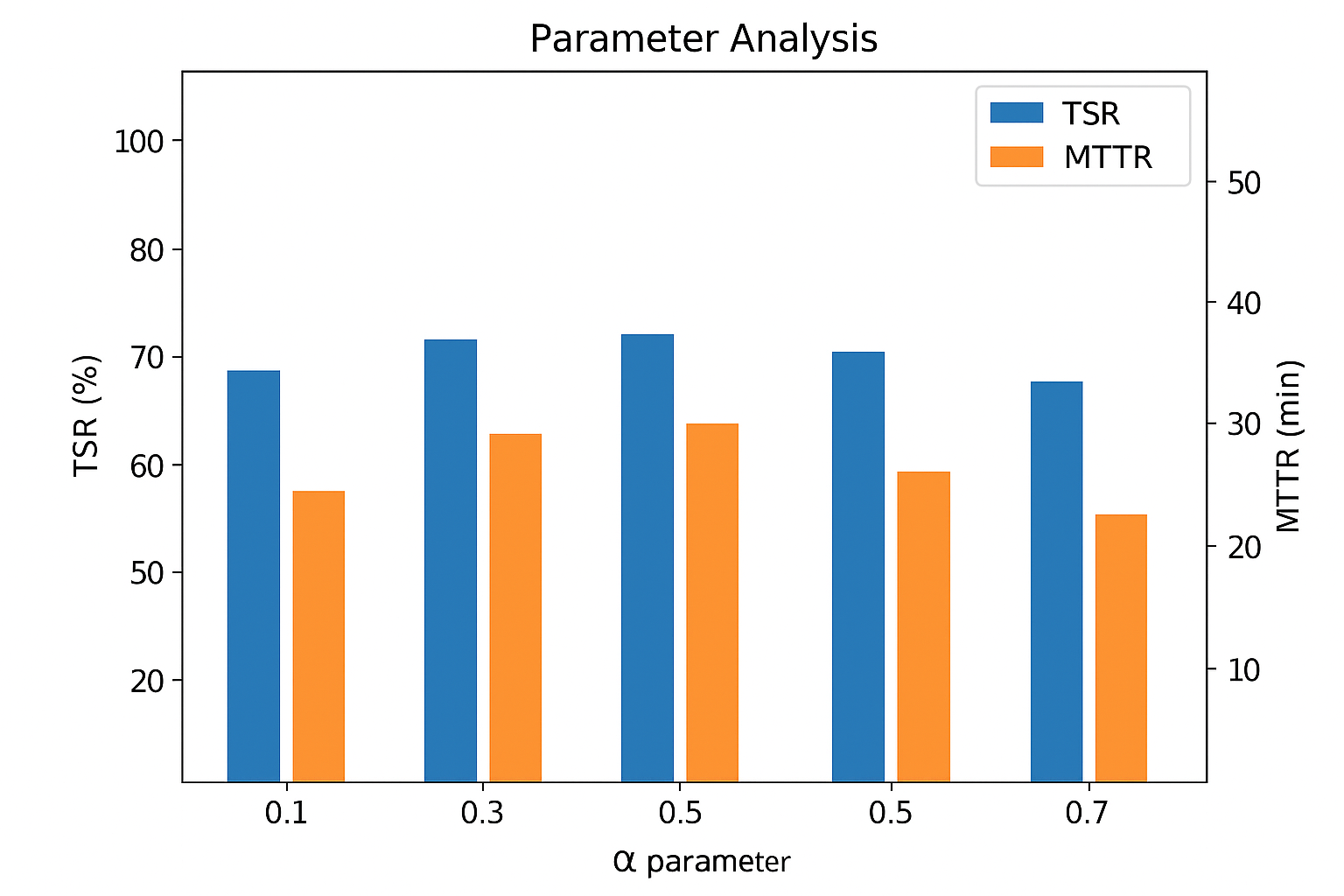}
\caption{Parameter sensitivity of AOI. The curves show that performance improves as the compressor receives enough local evidence, then saturates once larger windows add mostly redundant context.}
\label{fig:window_size_analysis}
\end{figure}

\noindent\textbf{Compression Window Size.}
Table~\ref{tab:param_sensitivity_window} and Figure~\ref{fig:window_size_analysis} show that increasing the compression window size consistently improves TSR, CCR, and IPS up to a moderate range, after which gains become marginal.
In particular, a window size of 768 tokens achieves the best overall trade-off, reaching a CCR of \textbf{72.4\%} and an IPS of \textbf{92.8\%}, while maintaining strong TSR.
Larger windows (e.g., 1024 and 1536 tokens) provide only negligible improvements at increased computational cost, indicating diminishing returns beyond this point.

\noindent\textbf{Scheduling Strategy Parameters.}
We vary the exploration--exploitation balance factor $\lambda \in [0.2, 0.8]$ and observe that performance peaks at $\lambda = 0.35$, suggesting that moderate exploration is critical for effective decision scheduling.

\noindent\textbf{Memory Retention Policies.}
Retaining episodic memory for 72 hours improves TSR by \textbf{3.5\%} compared to 24-hour retention, while extending retention to 120 hours yields minimal additional gains with higher resource overhead.

\begin{table}[!ht]
\centering
\caption{Sensitivity to compression window size. Results are averaged over five runs. A 768-token window provides the best practical trade-off: it preserves high task success and information retention while avoiding the extra cost of larger windows.}
\label{tab:param_sensitivity_window}
\resizebox{\linewidth}{!}{%
\begin{tabular}{@{}cccc@{}}
\toprule
Window size (tokens) & TSR (\%) & CCR (\%) & IPS (\%) \\
\midrule
256  & $89.1 \pm 1.3$ & $58.7$ & $86.2$ \\
512  & $91.4 \pm 1.0$ & $66.0$ & $90.1$ \\
\textbf{768}  & $\mathbf{94.2 \pm 0.8}$ & $\mathbf{72.4}$ & $\mathbf{92.8}$ \\
1024 & $94.5 \pm 0.9$ & $74.0$ & $92.9$ \\
1536 & $94.6 \pm 1.0$ & $74.6$ & $92.7$ \\
\bottomrule
\end{tabular}
}
\end{table}

\subsection{Scalability Study}
\begin{figure}[!ht]
\centering
\includegraphics[width=\columnwidth]{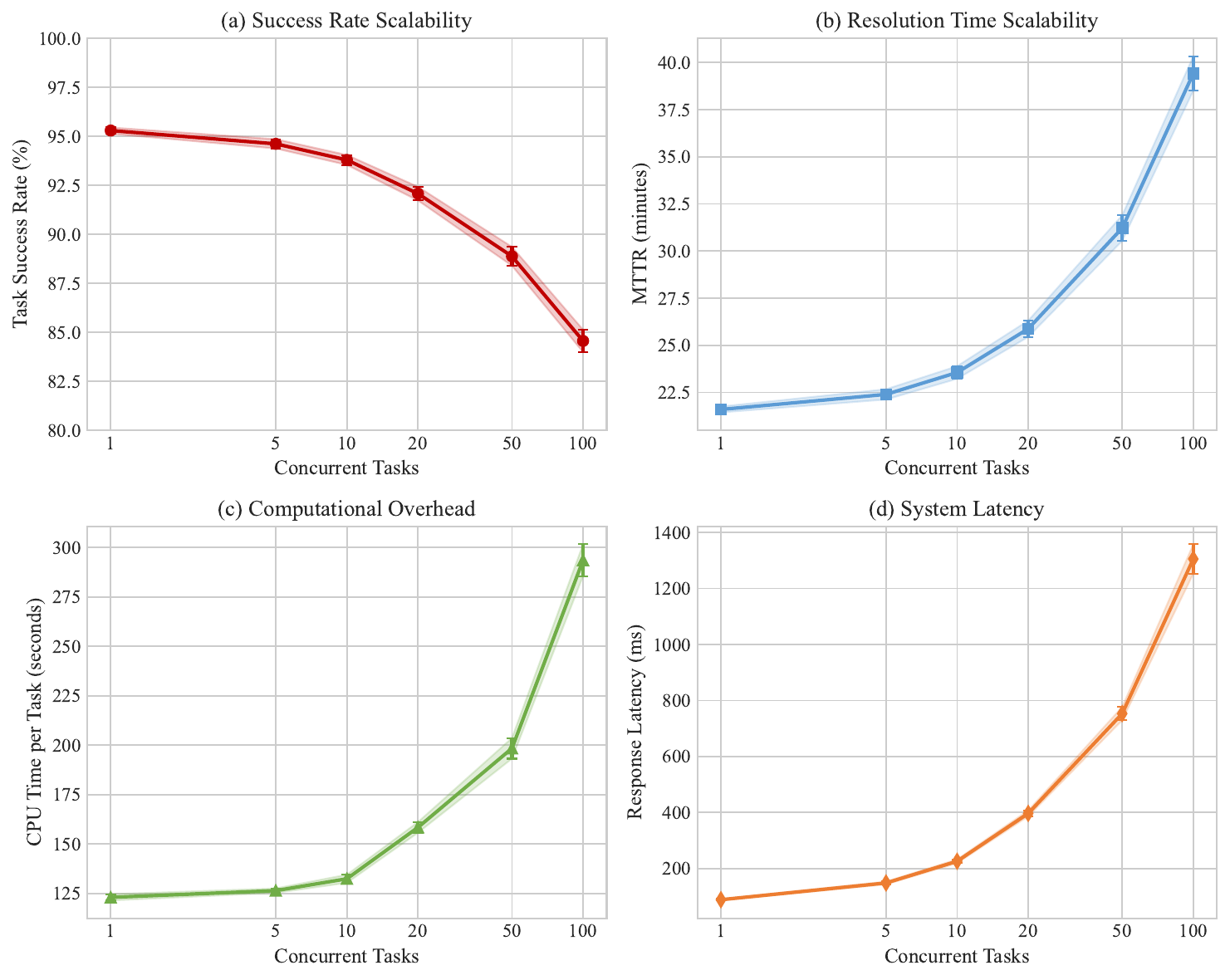}
\caption{Scalability profile under concurrent incidents. AOI maintains task success above 88\% up to 50 concurrent tasks, with resource use increasing smoothly as load grows.}
\label{fig:scalability_analysis}
\end{figure}

We evaluate the scalability of AOI under increasing levels of concurrency using the AIOpsLab simulation.
As shown in Figure~\ref{fig:scalability_analysis}, AOI maintains stable performance as the number of concurrent tasks grows, with task success rates remaining above 88\% up to 50 parallel tasks.
Table~\ref{tab:scalability} provides quantitative results, showing a gradual and controlled increase in MTTR as concurrency rises, while resource utilization efficiency (RUE) scales near-linearly with task load.
These results indicate that AOI avoids performance collapse under high concurrency and can support enterprise-scale deployments with sustained effectiveness.

\begin{table}[!ht]
\centering
\small
\caption{Scalability under increasing concurrent task load in AIOpsLab. Values are means over five runs. AOI degrades gradually rather than collapsing, maintaining high task success while MTTR and resource cost rise predictably.}
\label{tab:scalability}
\begin{tabular}{@{}cccc@{}}
\toprule
\textbf{Tasks} & \textbf{TSR} & \textbf{MTTR} & \textbf{RUE} \\
 & \textbf{(\%)} & \textbf{(min)} & \textbf{(CPU-s/task)} \\
\midrule
1   & $95.3 \pm 0.6$ & $21.6 \pm 0.9$ & $123$ \\
5   & $94.6 \pm 0.7$ & $22.3 \pm 1.4$ & $126$ \\
10  & $93.8 \pm 0.8$ & $23.5 \pm 1.3$ & $131$ \\
20  & $92.1 \pm 1.2$ & $25.8 \pm 1.4$ & $157$ \\
50  & $88.9 \pm 1.6$ & $31.2 \pm 2.7$ & $194$ \\
\bottomrule
\end{tabular}
\end{table}

\subsection{Qualitative Analysis and Case Studies}
\textbf{Case Study 1: Complex Cascading Failure.}  
In a simulated e-commerce infrastructure, AOI identified a cascading database connection failure within 4.3 minutes, automatically triggered dependency recovery, and fully restored service in 18.5 minutes. Competing systems required over 35 minutes and often failed to detect intermediate dependencies.

\noindent\textbf{Case Study 2: Configuration Drift Resolution.}  
In a Kubernetes-based microservice deployment, AOI detected a policy mismatch between ingress and service configurations. The Probe agent safely confirmed the issue, and the Executor executed a validated rollback plan with zero packet loss, resulting in overall downtime of less than 3 minutes.

\noindent\textbf{Failure Case Analysis.}  AOI exhibited limitations in scenarios involving extremely rapid failure propagation (less than 2 seconds), unseen hybrid fault types absent from the training corpus, and situations where GPT-based compression latency caused temporary decision delays under extreme load. These cases highlight areas for improvement and guide ongoing efforts toward online learning and real-time adaptive compression.

\begin{table}[!ht]
\centering
\small
\caption{Extended safety ablation. Results are five-run mean $\pm$ standard deviation. Removing safety verification or checkpointing harms recovery reliability even when the main agent loop remains intact, confirming that guarded execution contributes beyond context management alone.}
\label{tab:ablation_detailed}
\resizebox{\linewidth}{!}{%
\begin{tabular}{@{}lccc@{}}
\toprule
\textbf{Configuration} & \textbf{TSR} & \textbf{MTTR} & \textbf{CCR} \\
 & \textbf{(\%)} & \textbf{(min)} & \textbf{(\%)} \\
\midrule
AOI (Full) & $\mathbf{94.2 \pm 0.8}$ & $\mathbf{22.1 \pm 1.0}$ & $\mathbf{72.4}$ \\
w/o Context Compressor & $88.5 \pm 1.2$ & $29.8 \pm 1.5$ & N/A \\
w/o Dynamic Scheduling & $90.1 \pm 1.0$ & $26.7 \pm 1.3$ & $70.5$ \\
w/o Three-Layer Memory & $89.4 \pm 1.1$ & $27.5 \pm 1.4$ & $65.2$ \\
w/o Safety Verifier & $92.0 \pm 0.9$ & $24.3 \pm 1.2$ & $72.1$ \\
w/o Checkpointing & $91.1 \pm 1.0$ & $25.8 \pm 1.6$ & $71.9$ \\
Single-Agent Version & $84.6 \pm 1.4$ & $31.4 \pm 1.9$ & $60.3$ \\
\bottomrule
\end{tabular}
}
\end{table}

\begin{table*}[!ht]
\centering
\small
\caption{Cross-dataset summary of Task Success Rate (TSR) and Mean Time to Resolution (MTTR). Results are reported as mean $\pm$ standard deviation over five independent runs. AOI is best on every dataset, and the right panel highlights its margin over the strongest baseline.}
\label{tab:cross_dataset_summary}
\begin{tabular}{@{}llccclc@{}}
\toprule
Dataset & Method & TSR (\%) & MTTR (min) & & Dataset & AOI TSR / MTTR \\ 
\midrule
\multirow{5}{*}{HDFS} 
 & RES      & $62.4 \pm 2.1$ & $58.7 \pm 3.5$ & & HDFS (AOI)     & $\mathbf{92.1 \pm 1.1}$ / $\mathbf{22.9 \pm 1.2}$ \\
 & TAP      & $71.8 \pm 1.9$ & $44.9 \pm 2.8$ & & B-MAS (best)   & $84.7 \pm 1.4$ / $34.8 \pm 1.9$ \\
 & SA-LLM   & $78.9 \pm 1.6$ & $40.6 \pm 2.1$ & &  &  \\
 & B-MAS    & $84.7 \pm 1.4$ & $34.8 \pm 1.9$ & &  &  \\
 & \textbf{AOI} & $\mathbf{92.1 \pm 1.1}$ & $\mathbf{22.9 \pm 1.2}$ & &  &  \\
\midrule
\multirow{5}{*}{BGL} 
 & RES      & $66.1 \pm 2.4$ & $55.0 \pm 3.8$ & & BGL (AOI)      & $\mathbf{93.4 \pm 0.9}$ / $\mathbf{21.7 \pm 1.0}$ \\
 & TAP      & $74.0 \pm 2.0$ & $43.7 \pm 3.0$ & & B-MAS (best)   & $85.6 \pm 1.3$ / $33.2 \pm 2.0$ \\
 & SA-LLM   & $80.3 \pm 1.7$ & $39.1 \pm 2.4$ & &  &  \\
 & B-MAS    & $85.6 \pm 1.3$ & $33.2 \pm 2.0$ & &  &  \\
 & \textbf{AOI} & $\mathbf{93.4 \pm 0.9}$ & $\mathbf{21.7 \pm 1.0}$ & &  &  \\
\midrule
\multirow{5}{*}{OpenStack} 
 & RES      & $64.0 \pm 2.3$ & $56.9 \pm 3.6$ & & OpenStack (AOI) & $\mathbf{92.7 \pm 1.0}$ / $\mathbf{22.4 \pm 1.3}$ \\
 & TAP      & $73.6 \pm 2.1$ & $45.5 \pm 2.9$ & & B-MAS (best)   & $86.1 \pm 1.2$ / $34.1 \pm 1.8$ \\
 & SA-LLM   & $79.8 \pm 1.8$ & $39.8 \pm 2.5$ & &  &  \\
 & B-MAS    & $86.1 \pm 1.2$ & $34.1 \pm 1.8$ & &  &  \\
 & \textbf{AOI} & $\mathbf{92.7 \pm 1.0}$ & $\mathbf{22.4 \pm 1.3}$ & &  &  \\
\midrule
\multirow{5}{*}{AIOpsLab (sim)} 
 & RES      & $67.8 \pm 2.0$ & $54.3 \pm 3.3$ & & AIOpsLab (AOI) & $\mathbf{94.2 \pm 0.8}$ / $\mathbf{22.1 \pm 1.0}$ \\
 & TAP      & $75.1 \pm 1.8$ & $42.6 \pm 2.7$ & & B-MAS (best)   & $86.4 \pm 1.1$ / $33.7 \pm 1.7$ \\
 & SA-LLM   & $81.5 \pm 1.5$ & $38.2 \pm 2.2$ & &  &  \\
 & B-MAS    & $86.4 \pm 1.1$ & $33.7 \pm 1.7$ & &  &  \\
 & \textbf{AOI} & $\mathbf{94.2 \pm 0.8}$ & $\mathbf{22.1 \pm 1.0}$ & &  &  \\
\bottomrule
\end{tabular}
\end{table*}

\section{Discussion}\label{sec:discussion}

\subsection{Key Insights and Implications}
Our experiments suggest three broader lessons for autonomous operations.

\textbf{Context management is a first-order design problem.} The drop observed when context compression is removed shows that information overload is not merely an efficiency issue. Poor context selection changes the agent's decision state and can lead to slower or less reliable recovery. AOI's memory-aware compression helps preserve the evidence that matters while reducing the amount of raw telemetry passed through the reasoning loop.

\textbf{Agent specialization improves operational safety.} Separating the read-only Probe from the state-changing Executor gives the system a natural safety boundary. The Observer can request more evidence without granting modification privileges, while the Executor acts only after validation and checkpointing. This division is especially important in infrastructure settings, where an incorrect action can amplify an incident.

\textbf{Dynamic scheduling is necessary for long incidents.} Static policies either over-probe, delaying recovery, or execute too early, increasing risk. AOI's scheduler adjusts this balance as evidence accumulates, which explains the MTTR gains in both the main comparison and the ablation study.

\subsection{Advantages and Limitations}
The main advantage of AOI is that it treats autonomous operations as a closed-loop control problem rather than a single prediction task. The architecture scales naturally because each agent has a narrow responsibility; the safety mechanisms reduce the chance of destructive intervention; and the compressed memory state allows the system to carry useful context across long incident trajectories. At the same time, AOI has limitations. LLM-based compression introduces latency and external service dependency. Multi-agent coordination adds overhead relative to a single pipeline. Performance also depends on the coverage of the evaluation scenarios, and completely novel hybrid failures remain difficult when no comparable evidence has been observed before.

\subsection{Practical Deployment Considerations}
Organizations deploying AOI should treat it as an operational control layer rather than a drop-in alert classifier. The framework requires integration with observability systems, change-management policies, and rollback mechanisms. It also benefits from audit trails that record why the Observer scheduled a probe or execution step, what evidence the Probe collected, and which safety checks were applied before the Executor acted. In regulated environments, these traces are important for accountability and post-incident review.
\section{Conclusion}
We introduced AOI, a context-aware multi-agent framework for autonomous IT operations. AOI separates observation, safe probing, and guarded execution, and connects these agents through dynamic scheduling and hierarchical memory compression. Experiments across simulated and real-world operational scenarios show that this design improves task success, reduces MTTR, preserves critical context under strong compression, and maintains safety under concurrent load. The ablations further show that the gains come from the interaction of specialization, scheduling, memory, and safety rather than from a single component. More broadly, AOI suggests a path from anomaly-centric AIOps toward autonomous recovery systems that reason over context, act conservatively, and learn from long incident trajectories.

\section*{Limitations}
AOI still has several limitations. Its LLM-based compressor introduces latency and may depend on external model availability. The multi-agent design also requires additional orchestration and monitoring compared with a simpler detect-and-act pipeline. Although the evaluation covers diverse simulated and real-world scenarios, operational environments can contain failure modes that are absent from the benchmark distribution. Finally, the current framework focuses on technical recovery behavior and does not fully model organizational policies, approval workflows, or human escalation protocols.

\section{Future Work}\label{sec:future_work}
Several directions follow from AOI's results. First, future work should study privacy-preserving cross-organization learning. Operational incidents often repeat across organizations, but raw logs and remediation traces are sensitive. Federated or split-learning variants of AOI could allow sites to share compressed incident knowledge without exposing private telemetry.

Second, the context compressor can be made more domain-specialized. The current design uses a general LLM with operational instructions; a model trained or adapted on incident timelines, dependency graphs, and remediation traces may improve the trade-off between compression ratio, information preservation, and latency. Such a model could also expose uncertainty estimates so that the scheduler knows when compression may have removed important evidence.

Third, stronger formal safety guarantees remain an important direction. AOI already uses role separation, validation, checkpointing, and runtime monitoring, but production deployment would benefit from richer specifications that connect high-level operational policies to low-level execution constraints. Extending the formal model to cover rollback completeness, privilege boundaries, and multi-step dependency effects would make autonomous execution easier to audit.

Finally, future versions should place humans more explicitly in the control loop. Many incidents are safe for automatic remediation, but others require operator judgment, business-priority trade-offs, or regulatory approval. A promising direction is mixed-initiative operation, where AOI handles evidence collection and low-risk repairs while escalating uncertain or high-impact decisions with concise context summaries and actionable alternatives.
\bibliography{custom}

\appendix
\clearpage

\section{Implementation Details}
\label{app:implementation_details}

This appendix provides additional implementation details for the AOI framework, including the agent runtime, inter-agent communication, memory management, context compression, scheduling, and safety-control mechanisms.

\subsection{Agent Runtime and Communication}

Each AOI agent is implemented as an independent containerized Python service. The Observer, Probe, Executor, and Context Compressor are deployed as separate microservices and communicate through RESTful APIs implemented with FastAPI. This design isolates agent responsibilities and allows each module to scale independently under high task loads.

The Observer maintains the global task state and dispatches subtasks to the Probe and Executor. The Probe only exposes read-only diagnostic endpoints, while the Executor exposes state-changing endpoints protected by safety checks and checkpointing. All messages exchanged between agents follow a structured JSON schema:
\begin{equation}
\begin{aligned}
m &= (\texttt{task\_id}, \texttt{agent\_role}, \\ &\texttt{action\_type}, \texttt{payload}, \texttt{timestamp}, \texttt{status}).
\end{aligned}
\end{equation}
This schema ensures that task provenance, execution state, and rollback metadata can be tracked throughout the full operation lifecycle.

\subsection{Memory Backend}

The three-layer memory architecture is implemented using a Redis Cluster. Raw Context Storage keeps unprocessed logs, metrics, traces, and execution outputs with a default retention time of 24 hours. Task Queue Management stores structured subtasks, priorities, dependencies, and execution states. The Compressed Context Cache stores LLM-generated summaries and reusable incident-level knowledge with a default retention time of 7 days.

Each memory entry is indexed by incident identifier, task identifier, timestamp, and component name. This indexing strategy allows the Observer to retrieve both local task context and historical context from semantically related incidents. For compressed context retrieval, we use keyword filtering followed by embedding similarity ranking, where embeddings are computed from compressed summaries.

\subsection{Context Compression Implementation}

The Context Compressor applies a sliding-window compression strategy to long operational contexts. Given raw context $C_{\mathrm{raw}}$, we split it into overlapping windows with window size $w$ and overlap ratio $\rho=0.5$. Each window is summarized by an LLM using an operations-specific instruction that asks the model to preserve error codes, abnormal metrics, dependency relations, suspected causes, executed actions, and unresolved risks.

The compressor produces a window-level summary for each segment and then performs a second merge stage to remove duplicated content from overlapping regions. If the compressed context still exceeds the target budget, a secondary compression step is applied while keeping all critical fields. The final compressed context is stored as a structured record:
\begin{equation}
\begin{aligned}
C_{\mathrm{comp}} = (&\texttt{symptoms}, \texttt{evidence}, \\
&\texttt{suspected\_causes}, \texttt{actions}, \\
&\texttt{risks}, \texttt{open\_questions}).
\end{aligned}
\end{equation}
This format is designed to support both diagnosis and action planning.

\subsection{Dynamic Scheduling Implementation}

The Observer schedules subtasks according to priority, dependency constraints, information sufficiency, and operational risk. Each candidate action $a$ receives a scheduling score:
\begin{equation}
\begin{aligned}
\mathrm{Score}(a,t) &=
\lambda I(a,t) + (1-\lambda) G(a,t) \\
&- \mu R(a,t) - \nu C(a,t),
\end{aligned}
\end{equation}
where $I(a,t)$ denotes expected information gain, $G(a,t)$ denotes estimated progress toward the target state, $R(a,t)$ denotes operational risk, and $C(a,t)$ denotes resource cost. In our experiments, the exploration--exploitation balance is set to $\lambda=0.35$ unless otherwise stated. The risk and cost coefficients are selected on validation scenarios and then fixed for all test scenarios.

Probe actions are favored when information sufficiency is low or when the Executor requires fresh state observations before a modification. Executor actions are considered only when their prerequisites are satisfied and their risk scores are below the safety threshold.

\subsection{Safety Verification and Rollback}

The Executor uses a conservative execution protocol. Before any state-changing action, the system creates a checkpoint of the affected service configuration and verifies the proposed action against a rule-based safety verifier. The verifier rejects actions that involve destructive commands, missing dependencies, invalid service targets, or operations outside the authorized scope.

After each execution step, the system checks health metrics and service-level indicators. If a critical failure is detected, the Executor immediately triggers rollback to the latest checkpoint. The rollback metadata includes the checkpoint identifier, affected components, executed actions, and validation results. This design supports rapid recovery and auditability.

\subsection{Reproducibility Notes}

To reduce nondeterminism from LLM-based components, we use deterministic decoding where supported and fix all available random seeds. For API-based LLM calls, we cache requests and responses during each run to ensure that repeated analyses use identical compressor outputs. All baselines are executed under the same resource limits, task budgets, and fault-injection scenarios as AOI.

\section{Proofs and Mathematical Background}

\subsection{Proof of Theorem~\ref{thm:scheduling_convergence}}
\label{sec:proof_scheduling_convergence}

\begin{proof}
We first control the cumulative decrease of the potential. Taking total expectation in the drift inequality and summing from $t=1$ to $T$ gives
\begin{equation}
\begin{aligned}
&\mathbb{E}[\Phi(\sigma_{T+1})] - \Phi(\sigma_1) \\
\leq &-\eta \sum_{t=1}^T \mathbb{E}\|\nabla \Phi(\sigma_t)\|^2 + \frac{T \eta^2 G^2}{2} + \sum_{t=1}^T \beta_t.
\end{aligned}
\end{equation}
Since $\Phi(\sigma_{T+1}) \geq 0$, we obtain
\begin{equation}
\eta \sum_{t=1}^T \mathbb{E}\|\nabla \Phi(\sigma_t)\|^2 \leq \Phi(\sigma_1) + \frac{T \eta^2 G^2}{2} + \sum_{t=1}^T \beta_t.
\end{equation}
Dividing by $T \eta$ yields
\begin{equation}
\frac{1}{T} \sum_{t=1}^T \mathbb{E}\|\nabla \Phi(\sigma_t)\|^2 \leq \frac{\Phi(\sigma_1)}{\eta T} + \frac{\eta G^2}{2} + \frac{1}{\eta T} \sum_{t=1}^T \beta_t.
\end{equation}
Using the bound $\beta_t \leq c_0 \lambda / \sqrt{t}$, we have
\begin{equation}
\sum_{t=1}^T \beta_t \leq c_0 \lambda \sum_{t=1}^T \frac{1}{\sqrt{t}} \leq 2 c_0 \lambda \sqrt{T}.
\end{equation}
Therefore,
\begin{equation}
\frac{1}{T} \sum_{t=1}^T \mathbb{E}\|\nabla \Phi(\sigma_t)\|^2 \leq \frac{\Phi(\sigma_1)}{\eta T} + \frac{\eta G^2}{2} + \frac{2 c_0 \lambda}{\eta \sqrt{T}}.
\end{equation}
This shows that the scheduling dynamics approach an approximately stationary regime as $T$ grows.

Next, we bound the effect of exploration on decision quality. By the UCB-type exploration rule, with probability at least $1-\delta$,
\begin{equation}
\frac{R_T}{T} \leq C \sqrt{\frac{|\mathcal{A}| \log(|\mathcal{A}|/\delta)}{T}}.
\end{equation}
Hence, the average reward of the learned scheduling policy differs from that of the best static policy by at most the right-hand side above.

We then convert the potential and regret bounds into a completion-time bound. By assumption, the expected completion time is controlled by the time-averaged potential. Combining the potential descent estimate with the regret bound implies that the performance gap between $\pi_T$ and $\pi^\star_{\mathrm{static}}$ satisfies
\begin{equation}
\begin{aligned}
&J(\pi_T) - J(\pi^\star_{\mathrm{static}}) \\\leq& C_1 \frac{\Phi(\sigma_1)}{T} + C_2 \eta G^2 + C_3 \lambda + C_4 \sqrt{\frac{|\mathcal{A}| \log(|\mathcal{A}|/\delta)}{T}},
\end{aligned}
\end{equation}
for suitable constants $C_1,C_2,C_3,C_4 > 0$. The first term comes from the initialization error, the second from stochastic optimization noise, the third from persistent exploration, and the fourth from finite-time exploration regret.

Finally, choose $\eta$ and $\lambda$ so that
\begin{equation}
C_2 \eta G^2 + C_3 \lambda \leq \frac{\epsilon}{2}.
\end{equation}
Then choose $T$ large enough so that
\begin{equation}
C_1 \frac{\Phi(\sigma_1)}{T} + C_4 \sqrt{\frac{|\mathcal{A}| \log(|\mathcal{A}|/\delta)}{T}} \leq \frac{\epsilon}{2}.
\end{equation}
A sufficient condition is
\begin{equation}
T \geq \frac{4 C_4^2 |\mathcal{A}| \log(|\mathcal{A}|/\delta)}{\epsilon^2},
\end{equation}
with $T$ also large enough to absorb the initialization term. Under these choices, we obtain
\begin{equation}
J(\pi_T) \leq J(\pi^\star_{\mathrm{static}}) + \epsilon
\end{equation}
with probability at least $1-\delta$. This completes the proof.
\end{proof}

\subsection{Proof of Theorem~\ref{thm:compression_stability}}\label{sec:proof_compression_stability}

\begin{proof}
Define the Lyapunov function
\begin{equation}
V(\mathcal{C}) = \frac{1}{2}\|\mathcal{C} - \mathcal{C}^*\|_{\mathcal{H}}^2.
\end{equation}
Using the assumed contraction-with-error property, we obtain
\begin{equation}
\begin{aligned}
V(\mathcal{F}(\mathcal{C})) &= \frac{1}{2}\|\mathcal{F}(\mathcal{C}) - \mathcal{C}^*\|_{\mathcal{H}}^2 \\&\leq \frac{1}{2}\bigl(\alpha \|\mathcal{C} - \mathcal{C}^*\|_{\mathcal{H}} + \epsilon_{\mathrm{LLM}}\bigr)^2.
\end{aligned}
\end{equation}
Expanding the square gives
\begin{equation}
\begin{aligned}
V(\mathcal{F}(\mathcal{C}))
&\leq \frac{\alpha^2}{2}\|\mathcal{C} - \mathcal{C}^*\|_{\mathcal{H}}^2 \\
&\quad + \alpha \epsilon_{\mathrm{LLM}}\|\mathcal{C} - \mathcal{C}^*\|_{\mathcal{H}}
+ \frac{1}{2}\epsilon_{\mathrm{LLM}}^2.
\end{aligned}
\end{equation}
Subtracting
\begin{equation}
V(\mathcal{C}) = \frac{1}{2}\|\mathcal{C} - \mathcal{C}^*\|_{\mathcal{H}}^2
\end{equation}
from both sides yields
\begin{equation}
\begin{aligned}
V(\mathcal{F}(\mathcal{C})) - V(\mathcal{C}) &\leq -\frac{1-\alpha^2}{2}\|\mathcal{C} - \mathcal{C}^*\|_{\mathcal{H}}^2 \\&+ \alpha \epsilon_{\mathrm{LLM}}\|\mathcal{C} - \mathcal{C}^*\|_{\mathcal{H}} + \frac{1}{2}\epsilon_{\mathrm{LLM}}^2.
\end{aligned}
\end{equation}
This implies that the Lyapunov drift is strictly negative outside a neighborhood of $\mathcal{C}^*$ whose radius is controlled by $\epsilon_{\mathrm{LLM}}$. Therefore the iterates cannot diverge, and the state remains ultimately bounded.

To obtain the explicit bound, define
\begin{equation}
e_t = \|\mathcal{C}_t - \mathcal{C}^*\|_{\mathcal{H}}.
\end{equation}
Then the assumed contraction gives
\begin{equation}
e_{t+1} \leq \alpha e_t + \epsilon_{\mathrm{LLM}}.
\end{equation}
Unrolling this recursion yields
\begin{equation}
e_t \leq \alpha^t e_0 + \epsilon_{\mathrm{LLM}} \sum_{s=0}^{t-1} \alpha^s = \alpha^t e_0 + \epsilon_{\mathrm{LLM}} \frac{1-\alpha^t}{1-\alpha}.
\end{equation}
Taking the upper limit as $t \to \infty$, we obtain
\begin{equation}
\limsup_{t \to \infty} e_t \leq \frac{\epsilon_{\mathrm{LLM}}}{1-\alpha}.
\end{equation}
Hence the compression mechanism is practically stable. In the special case $\epsilon_{\mathrm{LLM}} = 0$, the recursion reduces to
\begin{equation}
e_{t+1} \leq \alpha e_t,
\end{equation}
which implies $e_t \to 0$ exponentially fast since $\alpha \in (0,1)$. Therefore $\mathcal{C}^*$ is asymptotically stable. This completes the proof.
\end{proof}

\subsection{Proof of Theorem~\ref{thm:info_preservation}}\label{thm:proof3}
\begin{proof}
Because $\mathcal{C}_{\mathrm{comp}}=\mathcal{F}_{w,\rho}(\mathcal{C})$ is generated from $\mathcal{C}$, we have the Markov relation
\begin{equation}
\mathcal{Y} \rightarrow \mathcal{C} \rightarrow \mathcal{C}_{\mathrm{comp}}.
\end{equation}
By the chain rule for mutual information,
\begin{equation}
I(\mathcal{Y};\mathcal{C},\mathcal{C}_{\mathrm{comp}})
=
I(\mathcal{Y};\mathcal{C}_{\mathrm{comp}})
+
I(\mathcal{Y};\mathcal{C}\mid \mathcal{C}_{\mathrm{comp}}).
\label{eq:chain_rule_1}
\end{equation}
Since $\mathcal{C}_{\mathrm{comp}}$ is a function of $\mathcal{C}$, adding $\mathcal{C}_{\mathrm{comp}}$ to $\mathcal{C}$ does not provide additional information beyond $\mathcal{C}$ itself. Hence,
\begin{equation}
I(\mathcal{Y};\mathcal{C},\mathcal{C}_{\mathrm{comp}})
=
I(\mathcal{Y};\mathcal{C}).
\label{eq:chain_rule_2}
\end{equation}
Combining Equations~\eqref{eq:chain_rule_1} and~\eqref{eq:chain_rule_2}, we obtain
\begin{equation}
I(\mathcal{Y};\mathcal{C})
=
I(\mathcal{Y};\mathcal{C}_{\mathrm{comp}})
+
I(\mathcal{Y};\mathcal{C}\mid \mathcal{C}_{\mathrm{comp}}).
\end{equation}
Rearranging gives
\begin{equation}
I(\mathcal{Y};\mathcal{C}_{\mathrm{comp}})
=
I(\mathcal{Y};\mathcal{C})
-
I(\mathcal{Y};\mathcal{C}\mid \mathcal{C}_{\mathrm{comp}}).
\end{equation}
Using the task-relevance loss condition in Equation~\eqref{eq:task_relevance_loss}, we have
\begin{equation}
I(\mathcal{Y};\mathcal{C}_{\mathrm{comp}})
\geq
I(\mathcal{Y};\mathcal{C})-\varepsilon_{w,\rho}.
\end{equation}
Substituting the window-overlap bound $\varepsilon_{w,\rho}\leq K/(w\rho)$ yields
\begin{equation}
I(\mathcal{Y};\mathcal{C}_{\mathrm{comp}})
\geq
I(\mathcal{Y};\mathcal{C})-\frac{K}{w\rho}.
\end{equation}
Finally, if $I(\mathcal{C};\mathcal{Y})>0$, dividing the additive bound by $I(\mathcal{C};\mathcal{Y})$ gives the relative form:
\begin{equation}
I(\mathcal{C}_{\mathrm{comp}};\mathcal{Y})
\geq
\left(1-\frac{K}{w\rho I(\mathcal{C};\mathcal{Y})}\right)
I(\mathcal{C};\mathcal{Y}).
\end{equation}
This completes the proof.
\end{proof}

\end{document}